# How to predict and avert economic crisis

Yong Tao[*]

School of Economics and Business Administration, Chongqing University, Chongqing 400044, China

**Abstract:** Our study shows that many firms would accumulate at zero output level (namely, Bankruptcy status) if a perfectly competitive market reaches full employment (namely, those people who should obtain employment have obtained employment). As a result, appearance of economic crisis is determined by two points; that is, (a). Stock market approaches perfect competition; (b). Society reaches full employment. The empirical research of these two points would lead to early warning of economic crisis. Moreover, it is a surprise that the state of economic crisis would be a feasible equilibrium within the framework of the Arrow-Debreu model. That means that we can not understand the origin of economic crisis within the framework of modern economics, for example, the general equilibrium theory.

Before proceeding to our subject, I need to introduce some backgrounds of our study.

Recently, we have engaged in an attempt (Tao 2010); that is, how to derive the macroeconomic behaviors by using microeconomic behaviors.

To this end, we need to focus on a long-run market; that is, firms run on extremely short time intervals ("firm-run") compared to the market ("market-run"). For a long-run competitive market, the main conclusions of microeconomics can be summarized as three points as follows (Tao 2010).

**(1). A competitive market is composed of a vast number of firms (agents) and diverse industries.**

**(2). Each firm always gains zero economic profit (normal profit) and hence can freely move from one industry to another.**

**(3). Supply creates its own demand (Say's law).**

Microeconomic behaviors of each individual have been included in above three points. Now we need to ask what a law of behavior the macroeconomic system (which is composed of all individuals) would obey. Next, we give a main point.

---

[*] E-mail address: taoyingyong2007@yahoo.com.cn

In a free economic system, every agent has fair chance to engage in various economic activities; that is, the probability of entering an arbitrary industry and gaining an arbitrary return for each agent is equal. From the concept of natural rights, we can easily understand this point. Based on this point, we give a definition of a macroeconomic equilibrium system as follow.

**(4). A macroeconomic equilibrium system corresponds to a situation where every agent would enter an industry (which is most probably entered by him) and would gain a return (which is most probable gained by him).**

It is well known that the behaviors of microscopic particle are very complex compared to the behaviors of people. Nevertheless, statistical regularities might emerge in the law of great number, for example, the law of ideal gases emerges from the chaotic motion of individual molecules. In fact, for a long-run competitive market, if the number of agents is enough large, the macroeconomic behaviors of the market (which is composed of all agents) would obey statistical regularities.

Using the definition (4) of a macroeconomic equilibrium system and microeconomic behaviors (1)-(3), we can find a statistical regularity as follow (Tao 2010).

**(5). Perfectly competitive market obeys Bose-Einstein statistics; purely monopolistic-competitive market obeys Boltzmann statistics**

Interestingly, there is a singularity in Bose-Einstein statistical equation; it would lead to instability of perfectly competitive market (Tao 2010). A concrete conclusion is as follow.

**(6). Many firms would accumulate at zero output level if a perfectly competitive market reaches full employment.**

Clearly, it is a state of economic crisis, which has puzzled human for near 300 years. Nevertheless, it is a surprise that the state of economic crisis would be a feasible equilibrium within the framework of the Arrow-Debreu model (Tao 2010). That means that we can not understand the origin of economic crisis within the framework of modern economics, for example, the general equilibrium theory.

Why is the stock market instable? That is because the stock market (more specifically, the bull market) would come close to perfect competition if it satisfies efficient markets hypothesis.

Interestingly, perfectly competitive market is just an extreme market; hence we could explain the reason why the economic crisis almost always starts from economy overheated extreme market, for example, the bull market. Nevertheless, it is careful to be noted that stock market crash leads to economic crisis if and only if the society reaches full employment.

Finally we attempt to use a simple example to understand the meaning of the conclusion (6).

**In a featureless snowfield, one needs a color target for orientation; otherwise, they shall get snow-blindness. Likewise, when people face many homogeneous products in a perfectly competitive market, they would become mad attribute to blindness. For example, when the stock market reaches the bull market, every stock price would rise. Then people will be confused (blind), since they find which stock is chosen for investment makes no difference in the long run. Therefore, many people would crazily unload their stocks; these behaviors would lead to stock market crash.**

Now, we start to ask how to predict and avert economic crisis.

The conclusion (6) has implied that appearance of economic crisis is determined by two points; that is,

**(7). Stock market approaches perfect competition.**

**(8). Society reaches full employment.**

If (7) and (8) are satisfied at the same time, then we need to be on the alert of the economic crisis. Clearly, to predict economic crisis, we need to pay close attention to the empirical research of the points (7) and (8). On the one hand, we have suggested a resolving index of investment $\omega$ as follow (Tao 2010), which approaches zero if the stock market approaches perfect competition.

**(9). We suppose there are $H$ stocks and $m$ investors, and the $j$th investor can identify $H_j$ stocks. Therefore, the resolving index of investment of stocks is denoted by**

$$\omega = \frac{\frac{H_1}{H} + \frac{H_2}{H} + ... + \frac{H_m}{H}}{m} = \frac{\sum_{j=1}^{m} H_j}{mH}.$$

On the other hand, to measure whether society reaches full employment, we can consider to measure whether the social marginal labor-capital return approaches zero.

Clearly, to avert economic crisis, government should encourage monopolistic competition in the long-run policy, for example, prolonging patent terms, increasing differentiated products, etc.